\documentclass[epj]{svjour}
\pdfoutput=1
\RequirePackage{snapshot} 

\usepackage{amsmath}
\usepackage{amssymb}
\usepackage{graphicx}
\usepackage{hyperref}
\usepackage{xspace}
\usepackage{wasysym} 
\usepackage{epstopdf}
\usepackage{booktabs}
\usepackage{hyperref}
\usepackage{algorithmicx}
\usepackage{algorithm}
\usepackage{algpseudocode}
\usepackage{cite}
\usepackage[caption=false]{subfig}

\AtBeginDocument{
\heavyrulewidth=.08em
\lightrulewidth=.05em
\cmidrulewidth=.03em
\belowrulesep=.65ex
\belowbottomsep=0pt
\aboverulesep=.4ex
\abovetopsep=0pt
\cmidrulesep=\doublerulesep
\cmidrulekern=.5em
\defaultaddspace=.5em
}
\newcommand{\norm}[1]{\left\lVert#1\right\rVert}
\newcommand{\GeV}{\;\mathrm{GeV}}

\usepackage{xcolor}

\begin{document}

\title{Lund jet images from generative and cycle-consistent
  adversarial networks}

\titlerunning{Lund jet images with GANs}

\newcommand{\OXaff}{Rudolf Peierls Centre for Theoretical Physics,
  University of Oxford,\\
  Clarendon Laboratory, Parks Road, Oxford OX1 3PU}

\newcommand{\MUaff}{TIF Lab, Dipartimento di Fisica,
  Universit\`a degli Studi di Milano and INFN Milan,\\
  Via Celoria 16, 20133, Milano, Italy}

\author{Stefano Carrazza\inst{1} \and Fr\'ed\'eric A. Dreyer\inst{2}}

\institute{\MUaff \and \OXaff}

\abstract{We introduce a generative model to simulate radiation
  patterns within a jet using the Lund jet plane.
  We show that using an appropriate neural network architecture with a
  stochastic generation of images, it is possible to construct a generative
  model which retrieves the underlying two-dimensional distribution to within a
  few percent.
  We compare our model with several alternative state-of-the-art generative
  techniques.
  Finally, we show how a mapping can be created between different categories of
  jets, and use this method to retroactively change simulation settings or the
  underlying process on an existing sample.
  These results provide a framework for significantly reducing
  simulation times through fast inference of the neural network as
  well as for data augmentation of physical measurements.
  \PACS{
    {12.38.-t}{Quantum chromodynamics} \and
    {07.05.Mh}{Neural networks, fuzzy logic, artificial intelligence}
  }
}

\date{Received: date / Accepted: date}

\maketitle

\section{Introduction}
\label{sec:introduction}

\begin{figure}
  \centering
  \includegraphics[width=0.45\textwidth]{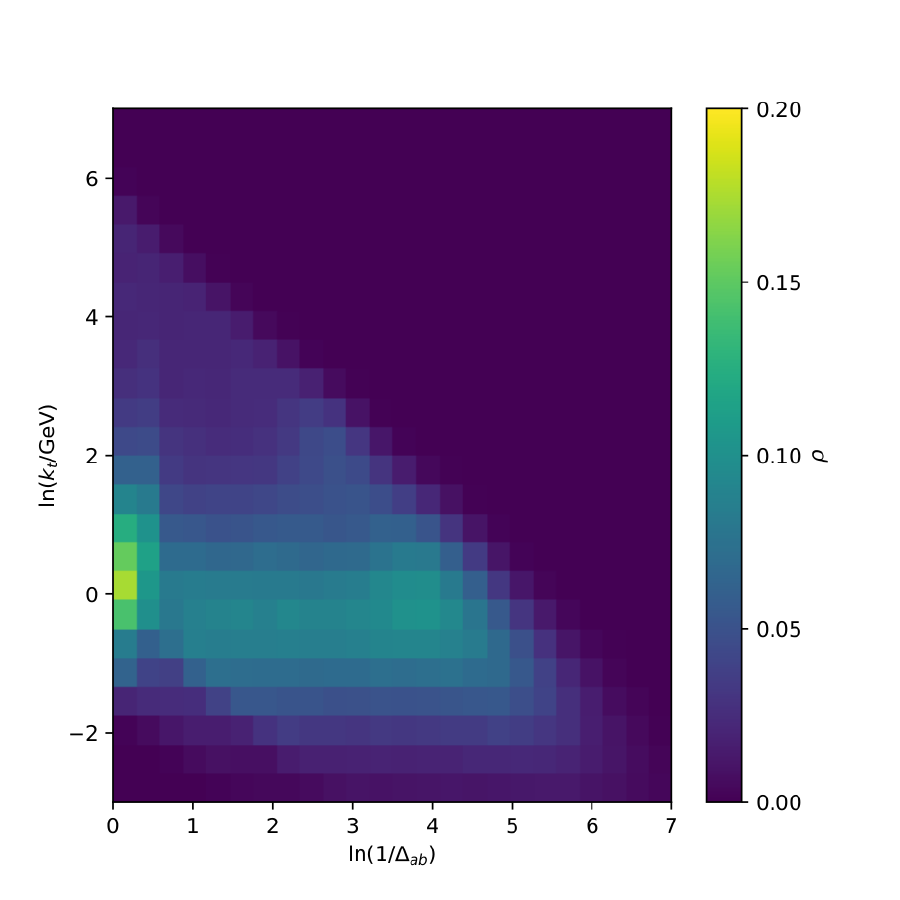}
  \caption{Average Lund jet plane density for QCD jets simulated with Pythia v8.223 and Delphes v3.4.1.}
  \label{fig:example_lund_plane}
\end{figure}

One of the most common objects emerging from hadron collisions at
particle colliders such as the Large Hadron Collider (LHC) are jets.
These are loosely interpreted as collimated bunches of energetic particles
arising from the interactions of quarks and gluons, the fundamental
constituents of the proton~\cite{Sterman:1977wj,Salam:2009jx}.
In practice, jets are usually defined through a sequential
recombination algorithm mapping final-state particle momenta to jet
momenta, with a free parameter $R$ defining the radius up to
which separate particles are clustered into a single
jet~\cite{Ellis:1993tq,Dokshitzer:1997in,Cacciari:2008gp}.

Because of the high energies involved in the collisions at the LHC,
heavy particles such as vector bosons or top quarks are frequently
produced with very large transverse momenta.
In this boosted regime, the decay products of these objects can become
so collimated that they are reconstructed as a single jet.
An active field of research is therefore dedicated to the theoretical
understanding of radiation patterns within jets, notably to
distinguish their physical origins and remove radiation unassociated
with the hard
process~\cite{Thaler:2008ju,Kaplan:2008ie,Ellis:2009su,Ellis:2009me,Plehn:2009rk,Thaler:2010tr,Larkoski:2013eya,Chien:2013kca,Cacciari:2014gra,Larkoski:2014gra,Moult:2016cvt,Dasgupta:2013ihk,Larkoski:2014wba,Komiske:2017ubm,Komiske:2017aww,Dreyer:2018tjj,Dreyer:2018nbf,Kasieczka:2019dbj,Carrazza:2019efs,Berta:2019hnj,Moreno:2019bmu}.
Furthermore, measurements of jet properties provide a unique
opportunity for accurate comparisons between theoretical predictions
and data, and can be used to tune simulation tools~\cite{Fischer:2014bja} or
extract physical constants~\cite{Bendavid:2018nar}.

In recent years, there has also been considerable interest in
applications of generative adversarial networks
(GAN)~\cite{goodfellow2014generative} and variational autoencoders
(VAE)~\cite{DBLP:journals/corr/KingmaW13} to particle physics, where
such generative models can be used to significantly reduce the
computing resources required to simulate realistic LHC
data~\cite{deOliveira:2017pjk,Paganini:2017hrr,Paganini:2017dwg,Otten:2019hhl,Musella:2018rdi,Datta:2018mwd,Cerri:2018anq,DiSipio:2019imz,Butter:2019cae,SHiP:2019gcl}.
In this paper, we introduce a generative model to create new samples
of the substructure of a jet from existing data.
We use the Lund jet plane~\cite{Dreyer:2018nbf}, shown in
figure~\ref{fig:example_lund_plane}, as a visual representation of the
clustering history of a jet.
This provides an efficient encoding of a jets radiation patterns and
can be directly measured experimentally~\cite{ATLAS:2019sol}.
The Lund jet image is used to train a Least Square GAN
(LSGAN)~\cite{DBLP:journals/corr/MaoLXLW16} to reproduce simulated
data within a few percent accuracy.
We compare a range of alternative generative methods, and show good agreement between the original jets generated with Pythia v8.223~\cite{Sjostrand:2014zea} using fast detector simulation with Delphes v3.4.1 particle flow~\cite{deFavereau:2013fsa} and samples provided by the different models~\cite{DBLP:conf/icml/ArjovskyCB17}.
Finally, we show how a cycle-consistent adversarial network
(CycleGAN)~\cite{CycleGAN2017} can be used to create mappings between
different categories of jets.
We apply this framework to retroactively change the parameters of the
parton shower on an event, adding non-perturbative effects to an
existing parton-level sample, and transforming quark and gluon jets to
a boosted $W$ sample.

These methods provide a systematic tool for data augmentation, as well
as reductions of simulation time and storage space by several orders
of magnitude, e.g. through a fast inference of the neural network
with hardware architectures such as GPUs and field-programmable gate
arrays (FPGA)~\cite{Duarte:2018ite}.
The code frameworks and data used in this work are available as
open-source and published material
in~\cite{frederic_dreyer_2019_3384921,frederic_dreyer_2019_3384919,gLund_data}\footnote{The codes are available
  at \url{https://github.com/JetsGame/gLund} and
  \url{https://github.com/JetsGame/CycleJet}}.

\section{Generating jets}
\label{sec:gen-jets}

In this article we will construct a generative model, which we call
gLund, to create new samples of radiation patterns of jets.
We first introduce the basis used to describe a jet as an
image, then construct a generative model which can be trained on these
objects.

\subsection{Encoding radiation patterns with Lund images}
\label{sec:lund-image}

To describe the radiation patterns of a jet, we will use the primary
Lund plane representation~\cite{Dreyer:2018nbf}, which can be projected
onto a two-dimensional image that serves as input to a neural network.

The Lund jet plane is constructed by reclustering a jet's constituents
with the Cambridge-Aachen (C/A)
algorithm\cite{Dokshitzer:1997in,Wobisch:1998wt}.
This algorithm sequentially recombines pairs of particles that have
the minimal $\Delta_{ij}^2=(y_i - y_j)^2 + (\phi_i - \phi_j)^2$ value,
where $y_i$ and $\phi_i$ are the rapidity and azimuth of particle $i$.

This clustering sequence can be used to construct an $n\times n$ pixel
image describing the radiation patterns of the initial jet.
We iterate in reverse through the clustering sequence, labelling the
momenta of the two branches of a declustering as $p_a$ and $p_b$,
ordered in transverse momentum such that $p_{t,a}>p_{t,b}$.
This procedure follows the harder branch $a$ and at each step we
activate the pixel on the image corresponding to the coordinates
$(\ln\Delta_{ab},\ln k_t)$, where $k_t=p_{t,b}\Delta_{ab}$ is the
transverse momentum of particle $b$ relative to $a$.%
\footnote{For simplicity, we consider only whether a pixel is on or
  off, instead of counting the number of hits as
  in~\cite{Dreyer:2018nbf}.
  While these two definitions are equivalent only for large image
  resolutions, this limitation can easily be overcome e.g. by
  considering a separate channel for each activation level.}

\subsection{Input data}
\label{sec:input-data}

The data sample used in this article consists of 500k jets, generated
using the dijet process in Pythia
v8.223.
Jets are clustered using the anti-$k_t$
algorithm~\cite{Cacciari:2008gp,Cacciari:2011ma} with radius $R=1.0$,
and are required to pass a selection cut, with transverse momentum
$p_t > 500$ GeV and rapidity $|y|<2.5$.
Unless specified otherwise, results use the
Delphes v3.4.1 fast detector simulation,
with the \verb|delphes_card_| \verb|CMS_NoFastJet.tcl| 
card to simulate both
detector effects and particle flow reconstruction.

The simulated jets are then converted to Lund images with $24\times24$
pixels each using the procedure described in
section~\ref{sec:lund-image}.
A pixel is set to one if there is a corresponding
$(\ln\Delta_{ab},\ln k_t)$ primary declustering sequence, otherwise it
is left at zero.

The full samples used in this article can be accessed
online~\cite{gLund_data}.

\subsection{Probabilistic generation of jets}
\label{sec:prob-generation}

\begin{figure}
  \centering
  \includegraphics[width=0.5\textwidth]{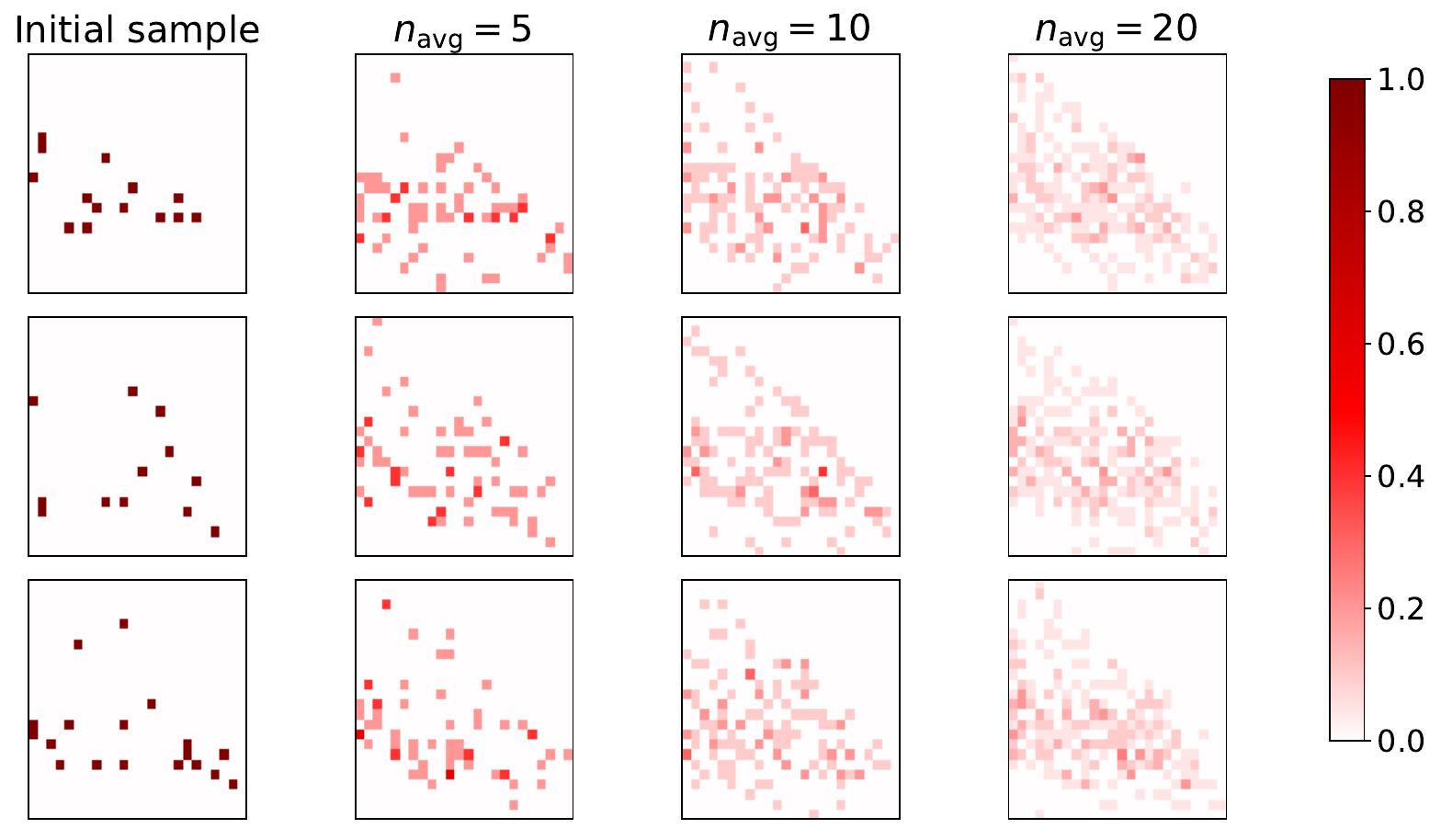}
  \caption{Sample input images after averaging with $n_\text{avg}=$1,
    5, 10 and 20.}
  \label{fig:navg-img}
\end{figure}

Generative adversarial networks~\cite{NIPS2014_5423} are one of the
most successful unsupervised learning methods.
They are constructed using both a generator $G$ and discriminator $D$,
which are competing against each other through a value function
$V(G,D)$.

In practice, we found improved performance when using a Least Square
Generative Adversarial Network
(LSGAN)~\cite{DBLP:journals/corr/MaoLXLW16}, a specific class of GAN
which uses a least squares loss function for the discriminator, and
has objective functions defined as

\begin{multline}
  \label{eq:gan-minD}
  \min_D V(D) = \frac12\mathbb{E}_{x\sim p_\text{data}}[(D(x)-b)^2]
  \\
  +\frac12\mathbb{E}_{z\sim p_z(z)}[(D(G(z))-a)^2]\,,
\end{multline}
\begin{equation}
  \label{eq:gan-minG}
  \min_G V(G) = \frac12\mathbb{E}_{z\sim p_z(z)}[(D(G(z))-c)^2]\,,
\end{equation}
where we defined $p_z(z)$ as a prior on input noise variables, and
$a$, $b$ and $c$ are the labels for the fake, real and presumed fake
data respectively.
Thus $D$ is trained in order to maximise the probability of correctly
distinguishing the training examples and the samples from $G$,
following equation~(\ref{eq:gan-minD}), while the latter is trained to
minimise equation~(\ref{eq:gan-minG}).
The generator's distribution $p_g$ optimises
equation~(\ref{eq:gan-minG}) when $p_g=p_\text{data}$, so that the
generator learns how to generate new samples from $z$.
The main advantage of the LSGAN over the original GAN framework is a more
stable training process, due to an absence of vanishing gradients.
In addition, we include a minibatch discrimination
layer~\cite{DBLP:journals/corr/SalimansGZCRC16} to avoid collapse of the
generator.

The LSGAN is trained on the full sample of QCD Lund jet images.
In order to overcome the limitation of GANs due to the sparse and
discrete nature of Lund images, we will use a probabilistic
interpretation of the Lund images to train the model.
To this end, we will first re-sample our initial data set into batches
of $n_\text{avg}$ and create a new set of 500k images, each consisting
of the average of $n_\text{avg}$ initial input images, as shown in
figure~\ref{fig:navg-img}.
These images can be reinterpreted as physical events through a random
sampling, where the pixel value is interpreted as the probability that
the pixel is activated.
The $n_\text{avg}$ value is a parameter of the model, with a large
value leading to increased variance in the generated images compared
to the reference sample, while for too low values the model performs
poorly due to the sparsity and discreteness of the data.
A further data preprocessing step before training the LSGAN consists in
rescaling the pixel intensities to be in the $[-1,1]$ range, and masking entries
outside of the kinematic limit of the Lund plane.
The images are then whitened using zero-phase components analysis (ZCA) whitening~\cite{BELL19973327}.

\subsection{gLund model results}
\label{sec:results}

The optimal choice of hyperparameters, both for the LSGAN model architecture and for the image preprocessing, is determined using the distributed asynchronous hyperparameter optimisation library \texttt{hyperopt}~\cite{Bergstra:2013:MSM:3042817.3042832}.

The performance of each setup is evaluated by a loss function which
compares the reference preprocessed Lund images to the artificial
images generated by the LSGAN model. We define the loss function as
\begin{equation}
  \mathcal{L}_h = I + 5 \cdot S
\end{equation}
where $I$ is the norm of the difference between the average of the images of the two samples and $S$ is the absolute difference in structural similarity~\cite{1284395} values between 5000 random pairs of reference samples, and reference and generated samples.

\begin{figure*}
  \centering
  \includegraphics[width=\textwidth]{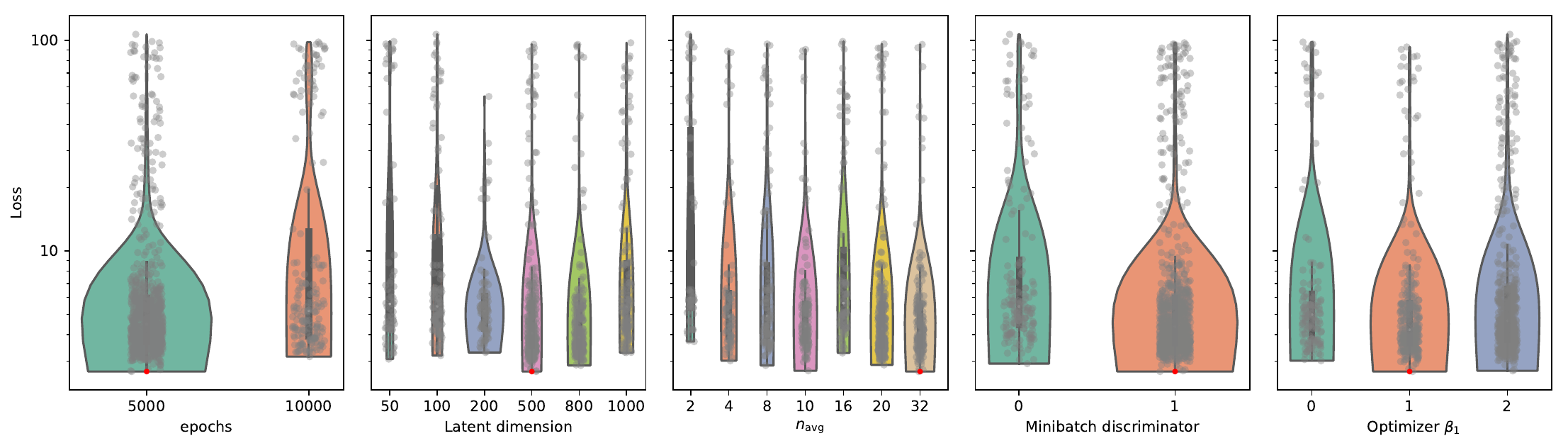}
  \includegraphics[width=\textwidth]{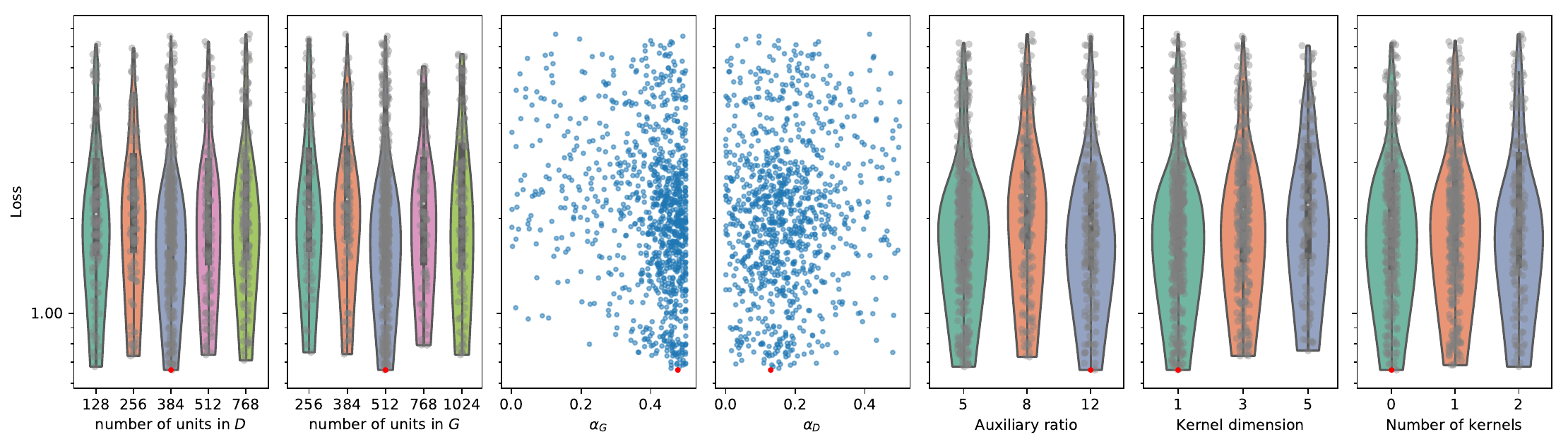}
  \caption{Hyperparameter scan results obtained with the hyperopt library. The first row shows the scan over image and optimiser related parameters while the second row plots correspond to the final architecture scan.}
  \label{fig:hyperopt}
\end{figure*}

\begin{figure}
  \centering
  \includegraphics[width=0.4\textwidth]{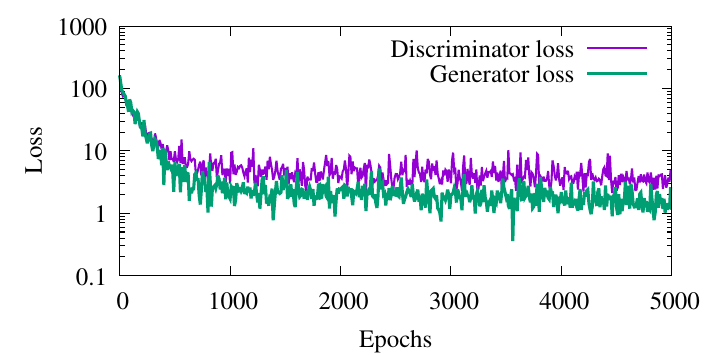}
  \caption{Loss of the LSGAN discriminator and generator throughout the training stage.}
  \label{fig:lsgan-loss}
\end{figure}

We perform 1000 iterations and select the one for which the loss
$\mathcal{L}_h$ is minimal. In figure~\ref{fig:hyperopt} we show some of the
results obtained with the \texttt{hyperopt} library through the Tree-structured
Parzen Estimator (TPE) algorithm. The LSGAN is constructed from a generator and
discriminator. The generator consists in three dense layers with 512, 1024 and
2048 units respectively using LeakyReLU~\cite{Maas13rectifiernonlinearities} activation
functions and batch normalisation layers, as well as a final layer matching the
output dimension and using a hyperbolic tangent activation function. The
discriminator is constructed from two dense layers with 768 and 384 units using
a LeakyReLU activation function, followed by another 24-dimensional dense layer
connected to a minibatch discrimination layer, with a final fully connected
layer with one-dimensional output.
The best parameters for this model are listed in table~\ref{tab:lsgan-parameters}.
The loss of the generator and discriminator networks of the LSGAN is
shown in figure~\ref{fig:lsgan-loss} as a function of training epochs.

\begin{figure*}
  \centering
  \includegraphics[width=\textwidth]{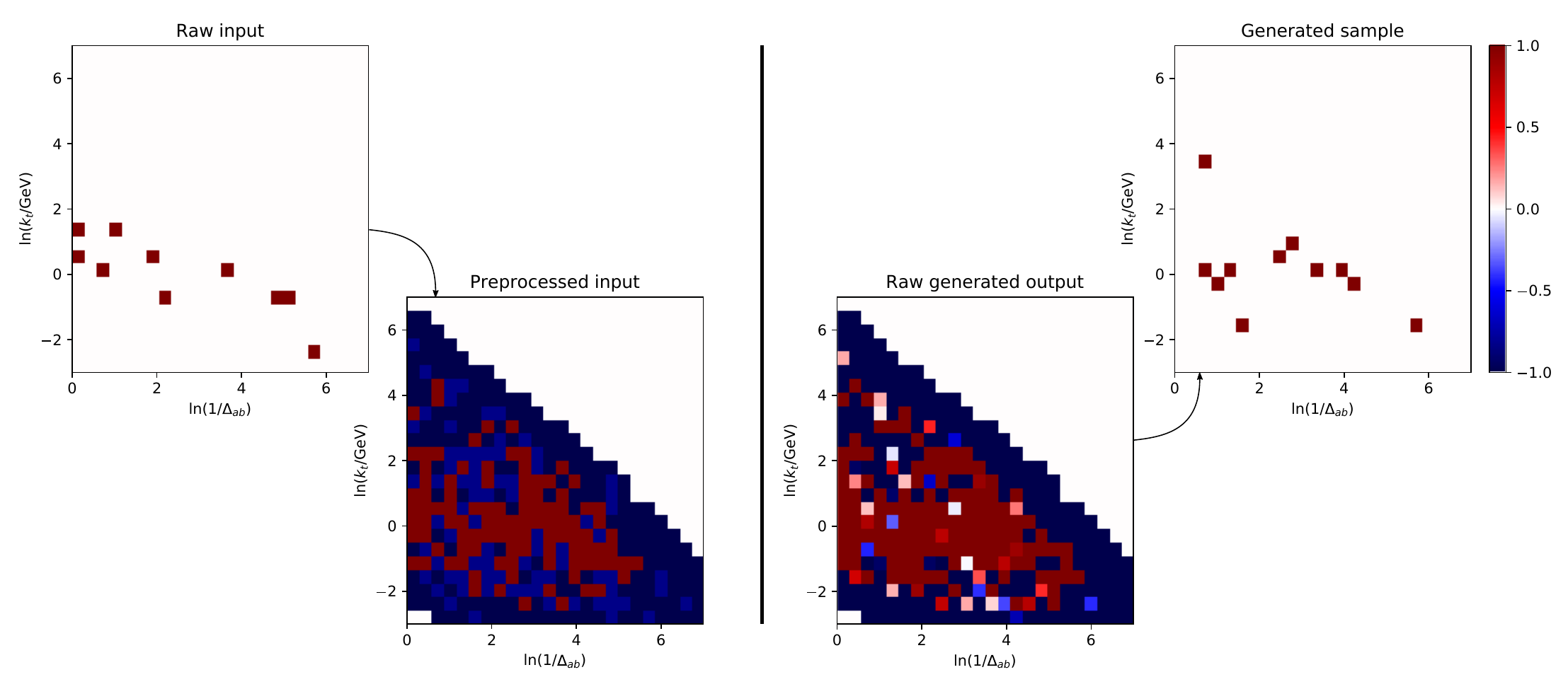}
  \caption{Left two figures: Sample input images before and after
    preprocessing. Right two: sample generated by the LSGAN and the
    corresponding Lund image.}
  \label{fig:events}
\end{figure*}

In figure~\ref{fig:events}, the first two images illustrate an example
of input image before and after preprocessing while the last two
images represent the raw output from the LSGAN model and the
corresponding sampled Lund image.

\begin{figure}
  \centering
  \includegraphics[width=0.5\textwidth]{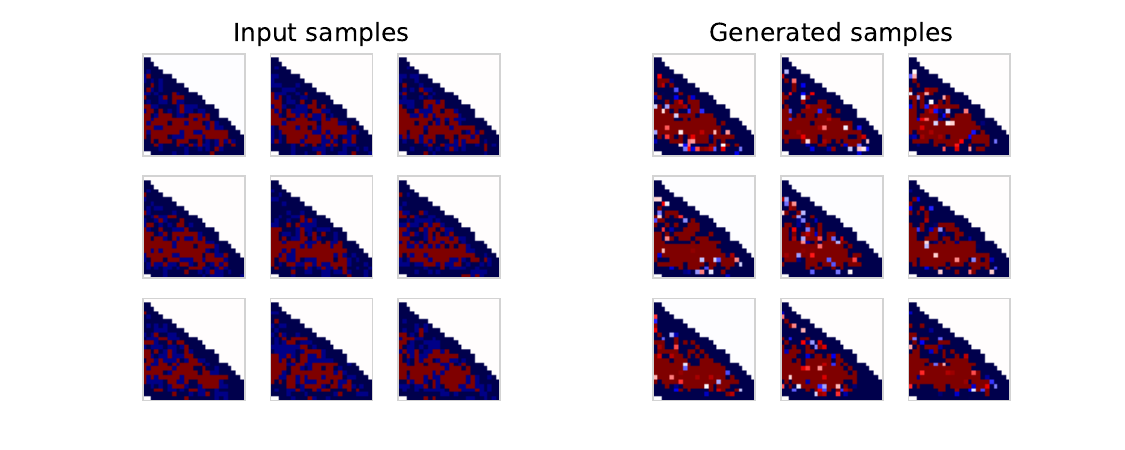}
  \caption{A random selection of preprocessed input images (left), and
    of images generated with the LSGAN model (right). Axes and colour
    schemes are identical to figure~\ref{fig:events}.}
  \label{fig:events2}
\end{figure}

\begin{figure*}
  \centering
  \subfloat[Reference]{\includegraphics[width=0.33\textwidth,page=4]{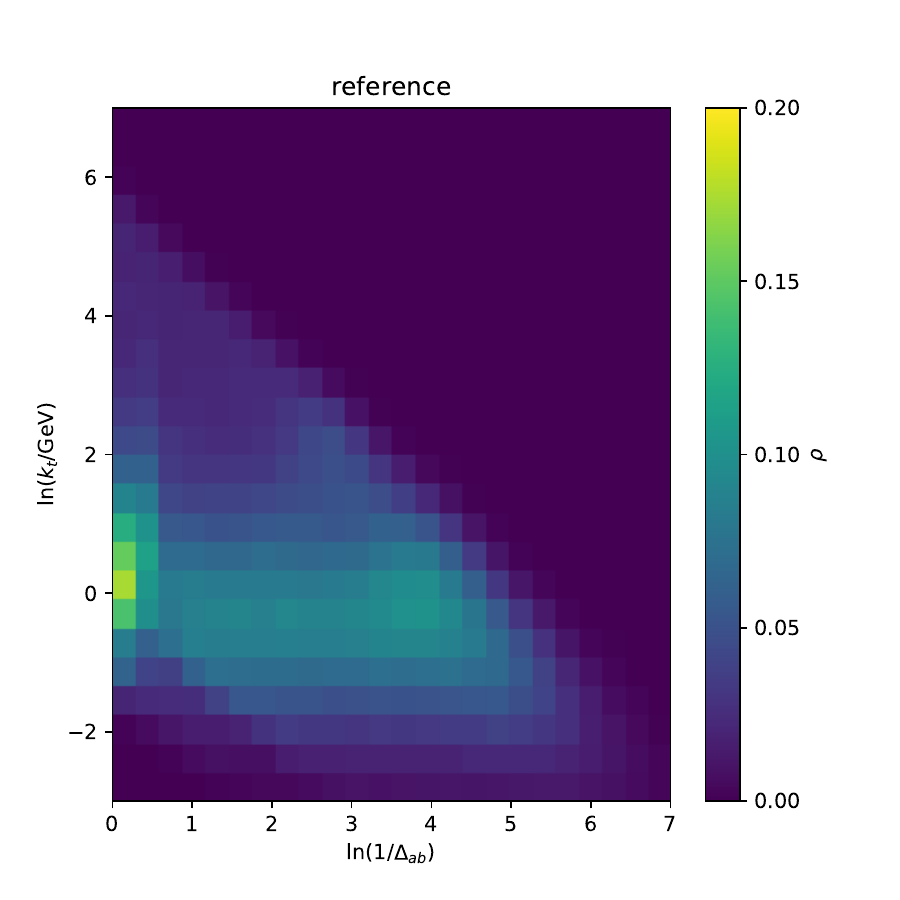}%
    \label{fig:ref_avg_img}}%
  \subfloat[Generated]{\includegraphics[width=0.33\textwidth,page=3]{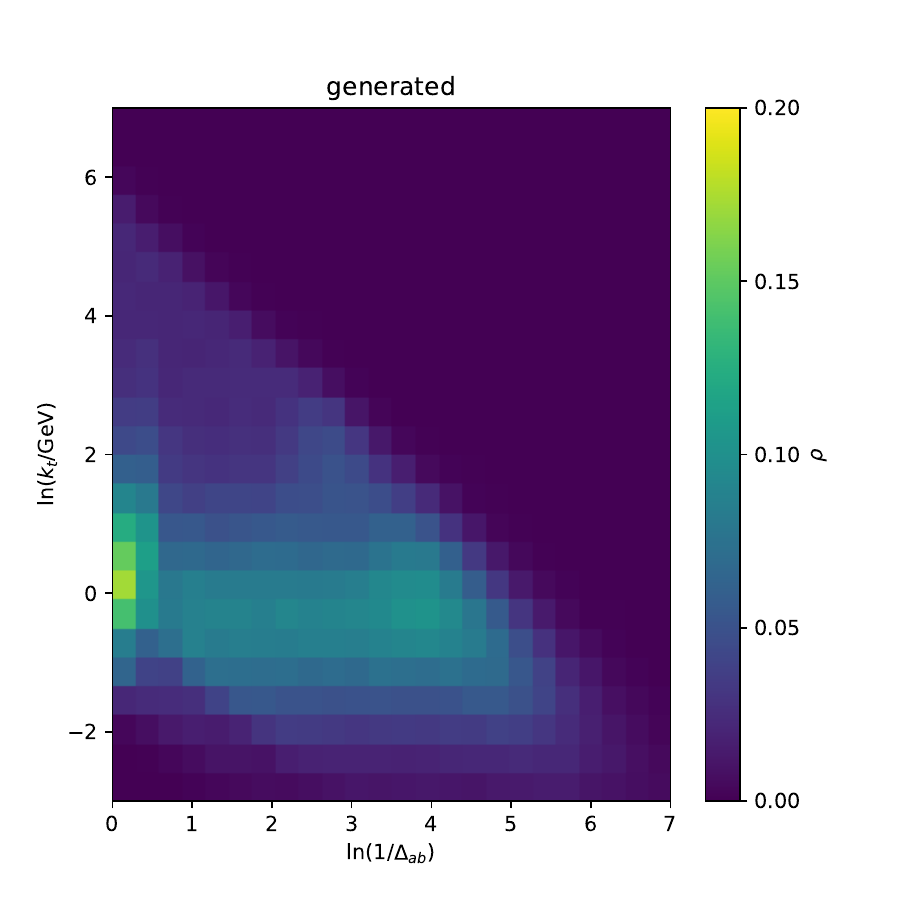}%
    \label{fig:gen_avg_img}}%
  \subfloat[Ratio generated/reference]{\includegraphics[width=0.33\textwidth,page=5]{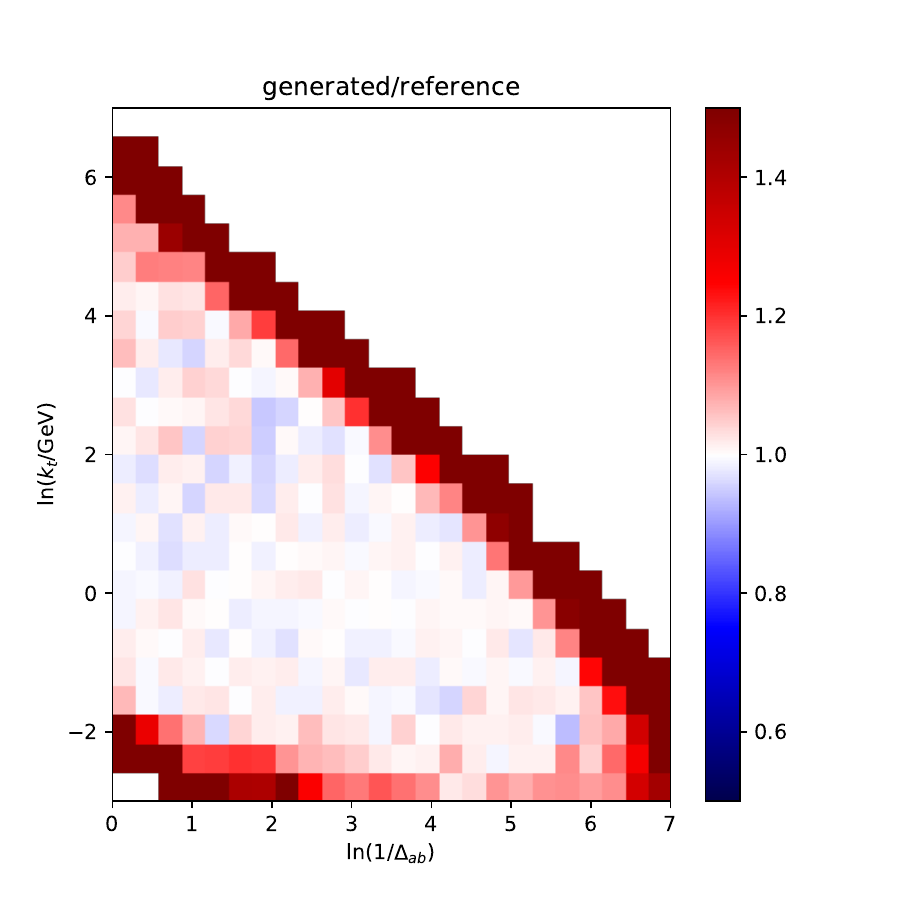}
    \label{fig:ratio_avg_img}}%
  \caption{Average Lund jet plane density for (a) the reference sample
    and (b) a data set generated by the gLund model. (c) shows the ratio
    between these two densities.}
  \label{fig:average_gen}
\end{figure*}

A selection of preprocessed input images and images generated with the LSGAN model are shown in figure~\ref{fig:events2}. The final averaged results for the Lund jet plane density are shown in figure~\ref{fig:average_gen} for the reference sample (left), the data set generated by the gLund model (centre) and the ratio between these two samples (right). We observe a good agreement between the reference and the artificial sample generated by the gLund model. The model is able to reproduce the underlying distribution to within a 3-5\% accuracy in the bulk region of the Lund image. Larger discrepancies are visible at the boundaries of the Lund image and are due the vanishing pixel intensities. In practice this model provides a new approach to reduce Monte Carlo simulation time for jet substructure applications as well as a framework for data augmentation.

\begin{table}
  \begin{center}
    \phantom{x}\medskip
    \begin{tabular}{ccc}
      \toprule
      \textbf{Parameters} && \textbf{Value}\\
      \midrule
      Architecture && LSGAN\\ [3pt]
      $D$ units && 384\\ [3pt]
      $G$ units && 512\\ [3pt]
      $\alpha_D$ && 0.129\\ [3pt]
      $\alpha_G$ && 0.477\\ [3pt]
      Aux ratio && 12\\ [3pt]
      Kernel dimension && 1\\ [3pt]
      Number of kernels && 2\\[3pt]
      Minibatch discriminator && Yes\\
      \midrule
      Epochs && 5000\\ [3pt]
      Batch size && 32\\ [3pt]
      Latent dimension && 500\\ [3pt]
      ZCA && Yes\\ [3pt]
      $n_{\rm avg}$ && 32\\ [3pt]
      Learning rate && $6.5 \cdot 10^{-5}$\\ [3pt]
      Decay $\beta_1$ && $8 \cdot 10^{-9}$\\ [3pt]
      Optimiser && Adagrad\\
      \bottomrule
    \end{tabular}
    \caption{Final parameters for the gLund model.}
    \label{tab:lsgan-parameters}
  \end{center}
\end{table}

\subsection{Comparisons with alternative methods}
\label{sec:comparisons}

Let us now quantify the quality of the model described in
section~\ref{sec:prob-generation} more concretely.
As alternatives, we consider a variational autoencoder
(VAE)~\cite{DBLP:journals/corr/KingmaW13,DBLP:journals/corr/HigginsMGPUBML16,DBLP:journals/corr/abs-1804-03599}
and a Wasserstein
GAN~\cite{DBLP:conf/icml/ArjovskyCB17,DBLP:journals/corr/GulrajaniAADC17}.

A VAE is a latent variable model, with a probabilistic encoder
$q_\phi(z|x)$, and a probabilistic decoder $p_\theta(x|z)$ to map a
representation from a prior distribution $p_\theta(z)$.
The algorithm learns the marginal likelihood of the data in this
generative process, which corresponds to maximising
\begin{equation}
  \label{eq:vaeloss}
  \mathcal{L}(\theta,\phi) = \mathbb{E}_{q_\phi(z|x)}[\log p_\theta(x|z)]
  - \beta D_{\text{KL}}(q_\phi(z|x)|| p(z))\,,
\end{equation}
where $\beta$ is an adjustable hyperparameter controlling the
disentanglement of the latent representation $z$.
In our implementation, we will set $\beta=1$, which corresponds to the
original VAE framework.

During the training of the VAE, we use KL cost
annealing~\cite{DBLP:journals/corr/BowmanVVDJB15} to avoid a collapse
of the VAE output to the prior distribution.
This is a problem caused by the large value of the KL divergence term
in the early stages of training, which is mitigated by adding a
variable weight $w_\text{KL}$ to the KL term in the cost function,
expressed as
\begin{equation}
  \label{eq:kl-anneal}
  w_\text{KL}(n_\text{step}) = \min(1,0.25\cdot1.05^{n_\text{step}})\,.
\end{equation}

Finally, we will also consider a Wasserstein GAN with gradient penalty
(WGAN-GP).
WGANs~\cite{DBLP:conf/icml/ArjovskyCB17} use the Wasserstein distance
to construct the value function, but can suffer from undesirable
behaviour due to the critic weight clipping.
This can be mitigated through gradient penalty, where the norm of the
gradient of the critic is penalised with respect to its
input~\cite{DBLP:journals/corr/GulrajaniAADC17}.

We determine the best hyperparameters for both of these models through
a \texttt{hyperopt} parameter sweep, which is summarised in
Appendix~\ref{app:models}.
To train these models using Lund images, we then use the same
preprocessing steps described in section~\ref{sec:prob-generation}.

To compare our three models, we consider two slices of fixed $k_t$ or
$\Delta_{ab}$ size, cutting along the Lund jet plane horizontally or
vertically respectively.

\begin{figure}
  \centering
  \includegraphics[width=0.45\textwidth]{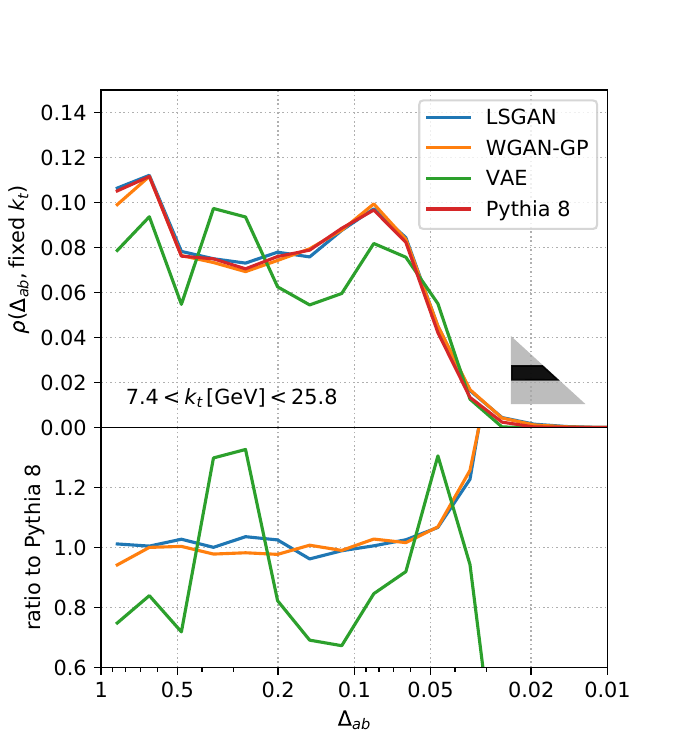}
  \caption{Slice of the Lund plane along $\Delta_{ab}$ with $7.4 \GeV < k_t < 25.8 \GeV$.}
  \label{fig:kt-slice}
\end{figure}

\begin{figure}
  \centering
  \includegraphics[width=0.45\textwidth]{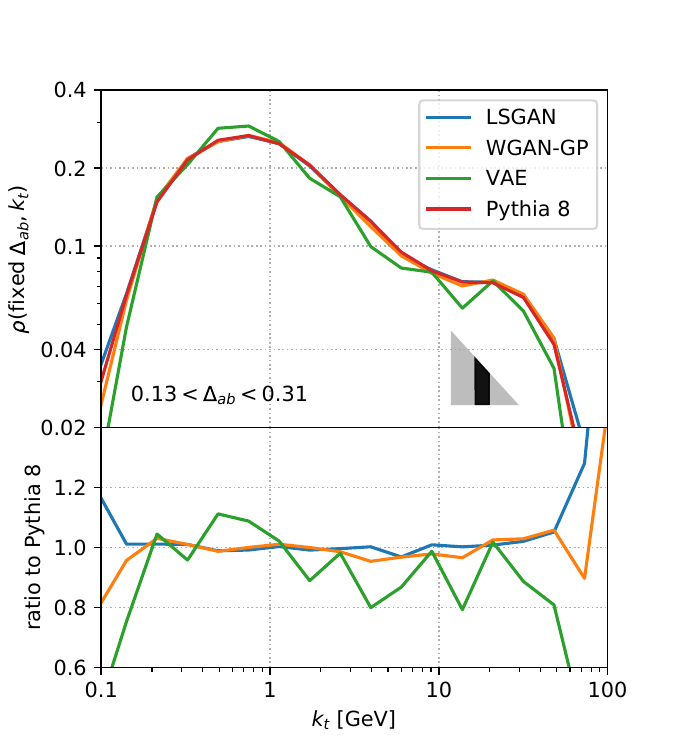}
  \caption{Slice of the Lund plane along $k_t$ with $0.13<\Delta_{ab}<0.31$.}
  \label{fig:delta-slice}
\end{figure}

In figure~\ref{fig:kt-slice}, we show the $k_t$ slice, with the
reference sample in red.
The lower panel gives the ratio of the different models to the
reference Pythia 8 curve, showing very good performance
for the LSGAN and WGAN-GP models, which are able to reproduce the data
within a few percent.
The VAE model also qualitatively reproduces the main features of the
underlying distribution, however we were unable to improve the
accuracy of the generated sample to more than $20\%$ without avoiding
the issue of posterior collapse.
The same observations can be made in figure~\ref{fig:delta-slice},
which shows the Lund plane density as a function of $k_t$, for a fixed
slice in $\Delta_{ab}$.

\begin{figure*}
  \centering
  \subfloat[]{\includegraphics[width=0.33\textwidth]{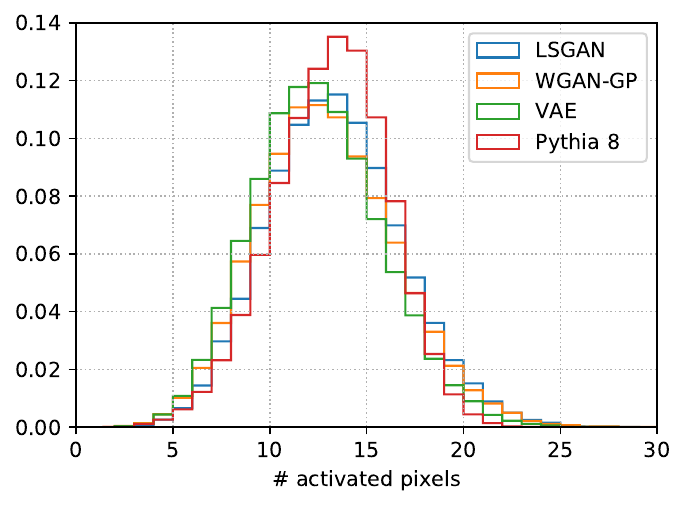}\label{fig:activation}}%
  \subfloat[]{\includegraphics[width=0.33\textwidth]{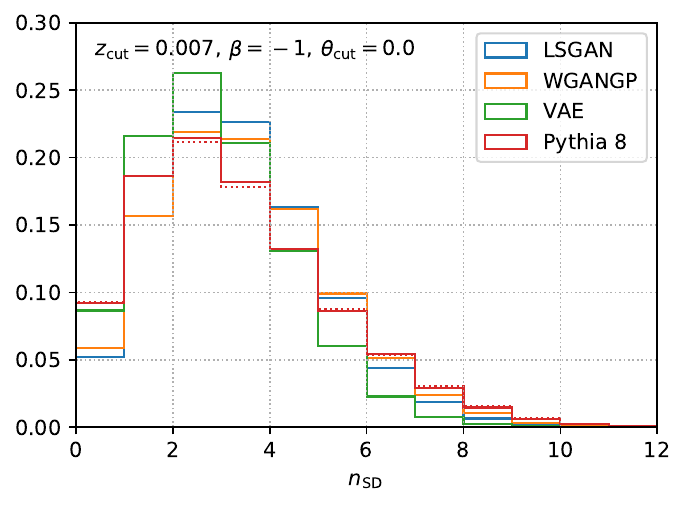}\label{fig:nsd}}%
  \subfloat[]{\includegraphics[width=0.32\textwidth]{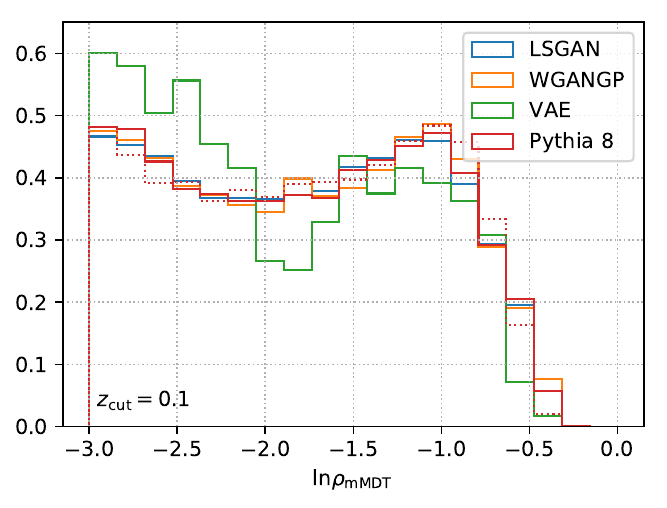}\label{fig:mass}}
  \caption{Distribution of (a) the number of activated pixels per
    image, (b) the reconstructed soft-drop multiplicity for
    $z_\text{cut}=0.007$, $\beta=-1$ and $\theta_\text{cut}=0$, and
    (c) the jet mass after applying the modified Mass Drop Tagger with
    $z_\text{cut}=0.1$.}
  \label{fig:observables}
\end{figure*}

In figure~\ref{fig:activation} we show the distribution of the number
of activated pixels per image for the reference sample generated with
Pythia 8 and the artificial images produced by the LSGAN, WGAN-GP and
VAE models. All models except the VAE model provide a good description
of the reference distribution.

We also use the Lund image to reconstruct the soft-drop
multiplicity~\cite{Frye:2017yrw}.
To this end, for a simpler correspondence between this observable and
the Lund image, we retrained the generative models using $\ln(z\Delta)$
as $y$-axis.
The soft-drop multiplicity can then be extracted from the final image,
and is shown in figure~\ref{fig:nsd} for each model using
$z_\text{cut}=0.007$ and $\beta=-1$.
The dashed lines indicate the true reference distribution, as
evaluated directly on the declustering sequence, and which differs
slightly from the reconstructed curve due to the finite pixel and
image size.

Finally, in figure~\ref{fig:mass}, we show the reconstructed mass of
the groomed jet using the modified Mass Drop
Tagger~\cite{Dasgupta:2013ihk} with $z_\text{cut}=0.1$, where we
approximate the mass as
\begin{equation}
  \label{eq:mass}
  \rho=\frac{m^2}{R^2 p_t^2}\simeq \max_i\Big[z^{(i)} \big(\Delta^{(i)}\big)^2\Big]\,,
\end{equation}
The dotted line shows the true mass distribution, evaluated with the
left-hand side of equation~(\ref{eq:mass}) on the groomed jet.
As in previous comparisons, we observe a very good agreement of the LSGAN
and WGAN-GP models with the reference sample.

We note that while the WGAN-GP model is able to accurately reproduce
the distribution of the training data, as discussed in
Appendix~\ref{app:models}, the individual images themselves can differ
quite notably from their real counterpart.
For this reason, our preferred model in this paper is the LSGAN-based
one.

\section{Reinterpreting events using domain mappings}
\label{sec:mappings}

In this section, we will introduce a novel application of domain
mappings to reinterpret existing event samples.
To this end, we implement a cycle-consistent adversarial network
(CycleGAN)~\cite{CycleGAN2017}, which is an unsupervised learning approach
to create translations between images from a source domain to a target
domain.

Using as input Lund images generated through different processes or
generator settings, one can use this technique to create mappings
between different types of jet.
As examples, we will consider a mapping from parton-level to
detector-level images, and a mapping from QCD images generated through
Pythia 8's dijet process, to hadronically decaying $W$ jets obtained
from $WW$ scattering.

\begin{figure}
  \centering
  \includegraphics[width=0.5\textwidth]{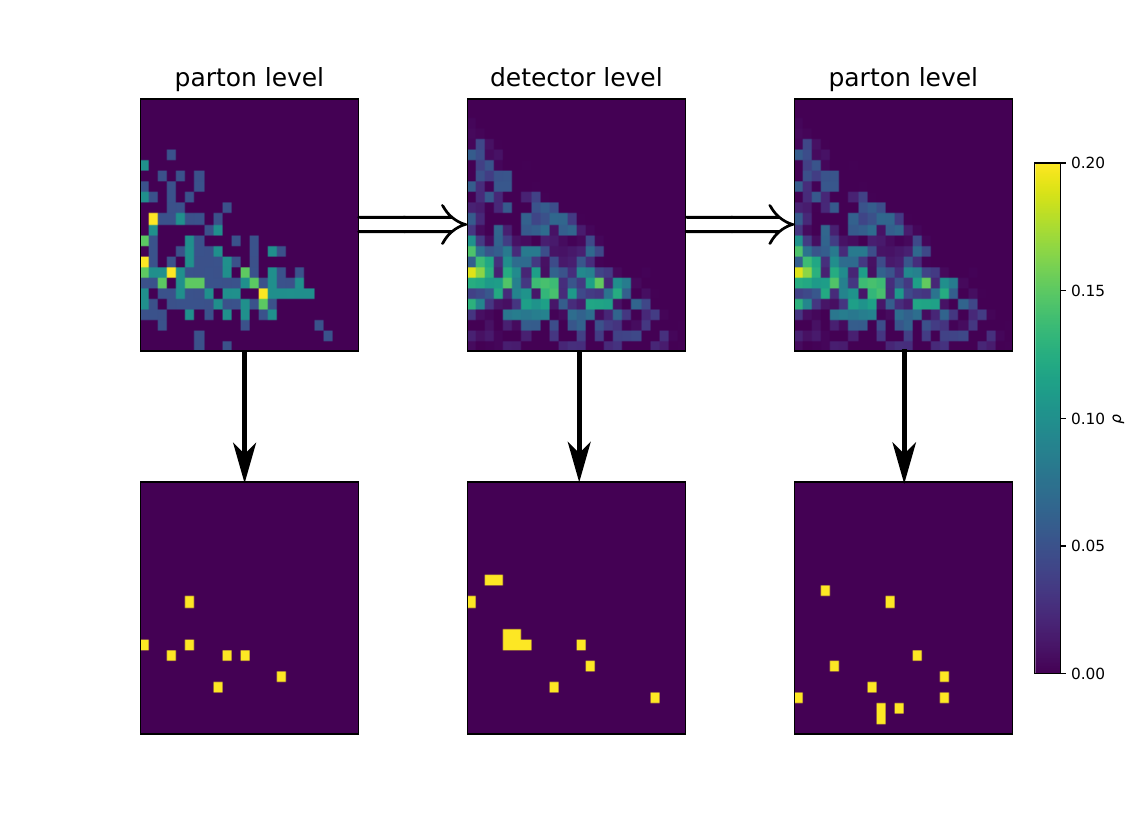}
  \caption{Top: transition from parton-level to delphes-level and back
    using CycleJet.
    Bottom: corresponding sampled event.}
  \label{fig:cyclejet0}
\end{figure}

The cycle obtained for a CycleGAN trained on parton and detector-level
images is shown in figure~\ref{fig:cyclejet0}, where an initial
parton-level Lund image is transformed to a detector-level one, before
being reverted again.
The sampled image is shown in the bottom row.

\subsection{CycleGANs and domain mappings}
\label{sec:cyclegan_intro}

A CycleGAN learns mapping functions between two domains $X$ and $Y$,
using as input training samples from both domains.
It creates an unpaired image-to-image translation by learning both a
mapping $G: X\rightarrow Y$ and an inverse mapping $F: Y\rightarrow X$
which observes a forward cycle consistency $x\in X\rightarrow G(x)\rightarrow F(G(x))\approx x$
as well as a backward cycle consistency $y\in Y\rightarrow F(y)\rightarrow G(F(y))\approx y$.
This behaviour is achieved through the implementation of a cycle
consistency loss
\begin{multline}
  \label{eq:cycle-loss}
  \mathcal{L}_\text{cyc}(G,F) = \mathbb{E}_{x\sim p_\text{data}(x)}[\norm{F(G(x)) - x}_1]\\
  +\mathbb{E}_{y\sim p_\text{data}(y)}[\norm{G(F(y)) - y}_1]\,,
\end{multline}

Additionally, the full objective includes also adversarial losses to
both mapping functions.
For the mapping function $G: X\rightarrow Y$ and its corresponding
discriminator $D_Y$, the objective is expressed as
\begin{multline}
  \label{eq:adv-loss}
  \mathcal{L}_\text{GAN}(G, D_Y, X, Y) = \mathbb{E}_{y\sim p_\text{data}(y)}[\log D_Y(y)]\\
  + \mathbb{E}_{x\sim p_\text{data}(x)}[\log (1 - D_Y(G(x)))]\,,
\end{multline}

such that $G$ is incentivized to generate images $G(x)$ that resemble
images from $Y$, while the discriminator $D_Y$ attempts to distinguish
between translated and original samples.

Thus, CycleGAN aims to find arguments solving
\begin{equation}
  \label{eq:cyclegan-soltn}
  G^*, F^* = \arg \min_{G,F} \max_{D_X, D_Y} \mathcal{L}(G,F,D_X, D_Y)\,,
\end{equation}
where $\mathcal{L}$ is the full objective, given by
\begin{multline}
  \label{eq:full-objective}
  \mathcal{L}(G,F,D_X, D_Y) = \mathcal{L}_\text{GAN}(G, D_Y, X, Y)\\
  + \mathcal{L}_\text{GAN}(F, D_X, Y, X) + \lambda\, \mathcal{L}_\text{cyc}(G,F)\,.
\end{multline}
Here $\lambda$ is parameter controlling the importance of the cycle
consistency loss.
We implemented a CycleGAN framework, labelled CycleJet, that can
be used to create mappings between two domains of Lund images.%
\footnote{CycleJet can also be used for similar practical purposes as
  DCTR~\cite{Andreassen:2019nnm}, albeit it is of course limited to
  the Lund image representation.}
By training a network on parton and detector-level images, this method
can thus be used to retroactively add non-perturbative and detector
effects to existing parton-level samples.
Similarly, one can train a model using images generated through two
different underlying processes, allowing for a mapping e.g. from QCD
jets to $W$ or top initiated jets.
%

\subsection{CycleJet model results}
\label{sec:cyclegan_results}

Following the pipeline presented in section~\ref{sec:results} we perform 1000
iterations of the hyperparameter scan using the \texttt{hyperopt} library and the loss function
\begin{equation}
  \mathcal{L}_h = || R_A - P_{B \rightarrow A} || + || R_B - P_{A \rightarrow B} ||
\end{equation}
where $A$ and $B$ indexes refer to the desired input and output samples respectively so $R_A$ and $R_B$ are the average reference images before the CycleGAN transformation while $P_{B \rightarrow A}$ and $P_{A \rightarrow B}$ correspond to the average image after the transformation.
Furthermore, for this model we noticed better results when preprocessing the
pixel intensities with the standardisation procedure of removing the mean and
scaling to unit variance, instead of a simpler rescaling in the [-1,1] range as done in section~\ref{sec:gen-jets}.

\begin{table}
  \begin{center}
    \phantom{x}\medskip
    \begin{tabular}{ccc}
      \toprule
      \textbf{Parameters} && \textbf{Value}\\
      \midrule
      $D$ filters && 32\\ [3pt]
      $G$ filters && 32\\ [3pt]
      $\lambda$ cycle && 10\\ [3pt]
      $\lambda$ identity factor && 0.2\\ [3pt]
      \midrule
      Epochs && 3\\ [3pt]
      Batch size && 128\\ [3pt]
      ZCA && Yes\\ [3pt]
      $n_{\rm avg}$ && 20\\ [3pt]
      Learning rate && $6.7 \cdot 10^{-3}$\\ [3pt]
      Decay $\beta_1$ && 0.7\\ [3pt]
      Optimiser && Adam\\
      \bottomrule
    \end{tabular}
    \caption{Final parameters for the CycleJet model.}
    \label{tab:cyclejet-parameters}
  \end{center}
\end{table}

The CycleJet model consists in two generators and two discriminators. The
generators consist in a down-sampling module with three two-dimensional
convolutional layers with 32, 64 and 128 filters respectively, and LeakyReLU
activation function and instance normalisation~\cite{DBLP:journals/corr/UlyanovVL16},
followed by an up-sampling with two two-dimensional convolutional layers with 64
and 32 filter. The last layer is a two-dimensional convolution with one filter
and hyperbolic tangent activation function. The discriminators take three
two-dimensional convolutional layers with 32, 64 and 128 filters and LeakyReLU
activation. The first convolutional layer has additionally an instance
normalisation layer and the final layer is a two-dimensional convolutional layer
with one filter.
The best parameters for the CycleJet model are shown in
table~\ref{tab:cyclejet-parameters}.

\begin{figure}
  \centering
  \includegraphics[width=0.5\textwidth,page=1]{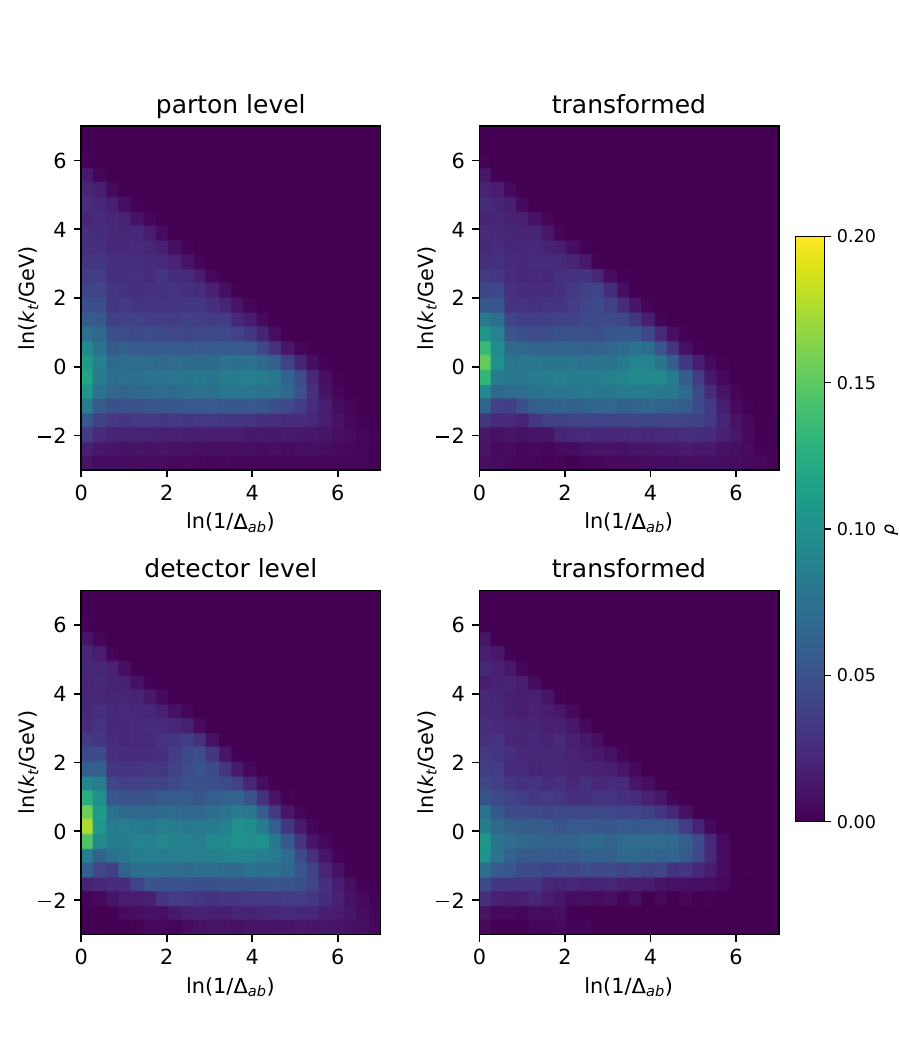}
  \caption{Top: average of the parton-level sample before (left) and after (right)
    applying the parton-to-detector mapping.
    Bottom: average of the delphes-level sample before (left) and after (right)
    applying the detector-to-parton mapping.}
  \label{fig:cyclejet1}
\end{figure}

\begin{figure}
  \centering
  \includegraphics[width=0.5\textwidth,page=1]{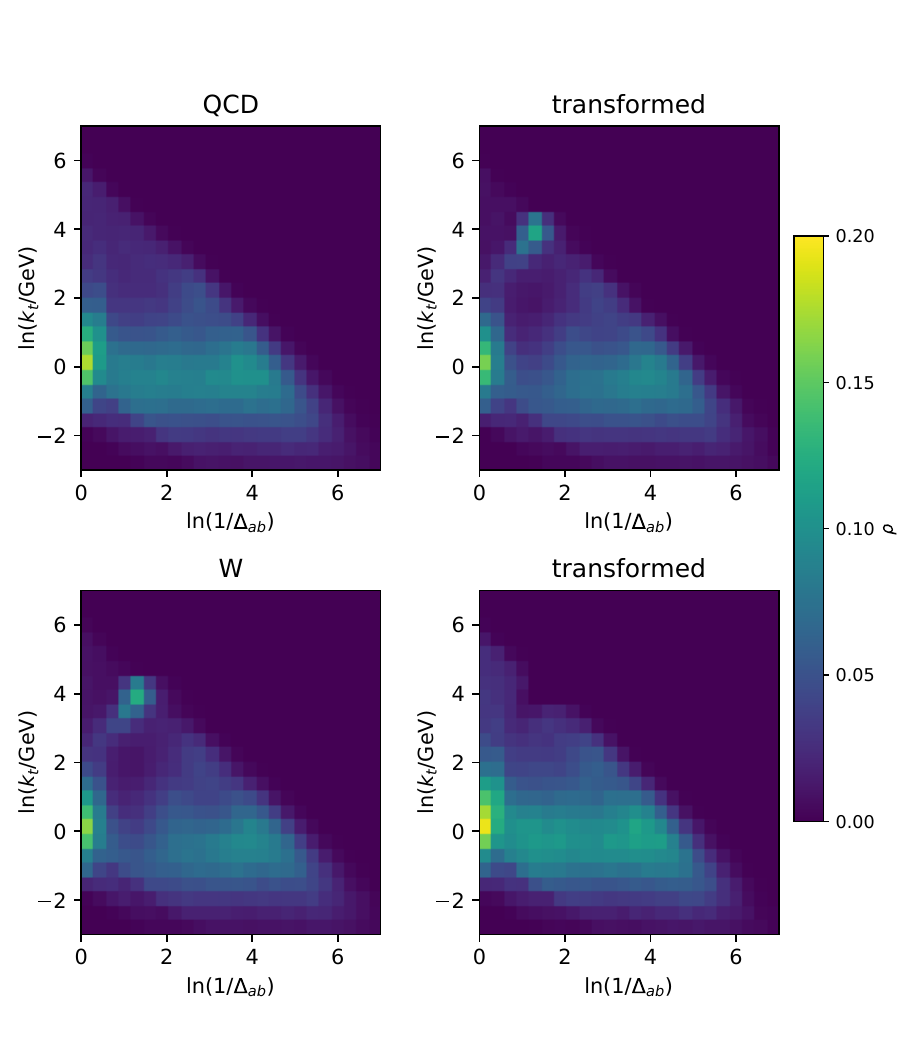}
  \caption{Top: average of the QCD sample before (left) and after (right)
    applying the QCD-to-$W$ mapping.
    Bottom: average of the $W$ sample before (left) and after (right)
    applying the $W$-to-QCD mapping.}
  \label{fig:cyclejet2}
\end{figure}

In the first row of figure~\ref{fig:cyclejet1} we show results for an initial
average parton-level sample before (left) and after (right) applying the
parton-to-detector mapping encoded by the CycleJet model, while in the second row of the same figure we perform the inverse operation by taking as input the average of the dephes-level sample before (left) and after (right) applying the CycleJet detector-to-parton mapping. This example shows clearly the possibility to add non perturbative and detector effects to a parton level simulation within good accuracy.
Similarly to the previous example, in figure~\ref{fig:cyclejet2} we present the
mapping between QCD-to-$W$ jets and vice-versa. Also in this case, the overall
quality of the mapping is reasonable and provides and interesting successful
test case for process remapping.

For both examples we observe a good level agreement for the respective
mappings, highlighting the possibility to use such an approach to save
CPU time for applying full detector simulations and non perturbative
effects to parton level events.
It is also possible to train the CycleJet model on Monte Carlo data and
apply the corresponding mapping to real data.

\section{Conclusions}
\label{sec:conclusions}

We have conducted a careful study of generative models applied to jet
substructure.

First, we trained a LSGAN model to generate new artificial samples of
detector level Lund jet images. With this, we observed agreement to
within a few percent accuracy in the bulk of the phase space with
respect to the reference data. This new approach provides an efficient
method for fast simulation of jet radiation patterns without requiring
the long runtime of full Monte Carlo event generators. Another
advantage consists in the possibility of this method to be applied to
real collider data to generate accurate physical samples, as well as
making it possible to avoid the necessity for large storage space by
generating realistic samples on-the-fly.

Secondly, a CycleGAN model was constructed to map different jet
configurations, allowing for the conversion of existing events. This
procedure can be used to change Monte Carlo parameters such as the
underlying process or the shower parameters. As examples we show how
to convert an existing sample of QCD jets into $W$ jets and
vice-versa, or how to add non perturbative and detector effects to a
parton level simulation. As for the LSGAN, this method can be used
to save CPU time by including full detector simulations and non
perturbative effects to parton level events. Additionally, one could
use CycleJet to transform real data using mappings trained on Monte
Carlo samples or apply them to samples generated through gLund.

To achieve the results presented in this paper we have implemented a rather
convolved preprocessing step which notably involved combining and resampling
multiple images. This procedure was necessary to achieve accurate distributions
but comes with the drawback of loosing information on correlations between
emissions at wide angular and transverse momentum separation. Therefore, it is
difficult to evaluate or improve the formal logarithmic accuracy of the
generated samples. This limitation could be circumvented with an end-to-end GAN
architecture more suited to sparse images. We leave a more detailed study of
this for future work.
The full code and the pretrained models presented in this paper are
available
in~\cite{frederic_dreyer_2019_3384921,frederic_dreyer_2019_3384919}.

\section*{Acknowledgments}
We thank Sydney Otten for discussions on $\beta$-VAEs.
We also acknowledge the NVIDIA Corporation for the donation of a Titan
Xp GPU used for this research.
F.D.\ is supported by the Science and Technology Facilities Council
(STFC) under grant ST/P000770/1. S.C.\ is supported by the European
Research Council under the European Union's Horizon 2020 research and
innovation Programme (grant agreement number 740006).

\appendix
\section{VAE and WGAN-GP models}
\label{app:models}

In this appendix we present the final parameters as well as generated event
samples for the VAE and WGAN-GP models used in section~\ref{sec:comparisons}.
These models are obtained after applying the \texttt{hyperopt} procedure
described in section~\ref{sec:results}.

\begin{table}
  \begin{center}
    \phantom{x}\medskip
    \begin{tabular}{ccc}
      \toprule
      \textbf{Parameters} && \textbf{Value}\\
      \midrule
      Intermediate dimension && 384\\ [3pt]
      KL annealing rate && 0.25\\ [3pt]
      KL annealing factor && 1.05\\ [3pt]
      Minibatch discriminator && No\\
      \midrule
      Epochs && 50\\ [3pt]
      Batch size && 32\\ [3pt]
      Latent dimension && 1000\\ [3pt]
      ZCA && Yes\\ [3pt]
      $n_{\rm avg}$ && 32\\ [3pt]
      Learning rate && $4.2 \cdot 10^{-4}$\\ [3pt]
      Decay $\beta_1$ && 0.9\\ [3pt]
      Optimiser && Adam\\
      \bottomrule
    \end{tabular}
    \caption{Final parameters for the VAE model.}
    \label{tab:vae-parameters}
  \end{center}
\end{table}

\begin{figure}
  \centering
  \includegraphics[width=0.5\textwidth]{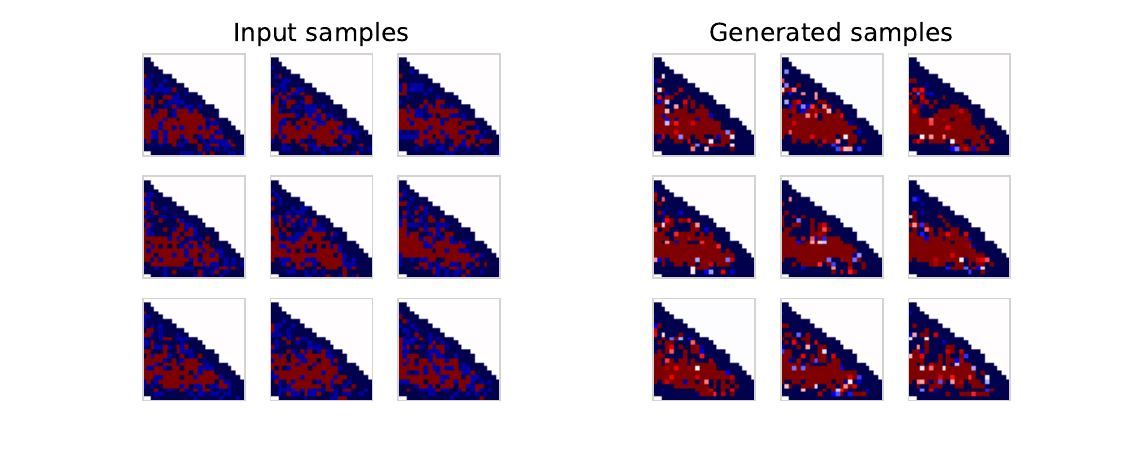}
  \caption{A random selection of preprocessed input images (left), and of
    images generated with the VAE model (right). Axis and colour schemes are the same of figure~\ref{fig:events}.}
  \label{fig:events-vae}
\end{figure}

The VAE encoder consists of a dense layer with 384 units with ReLU activation
function connected to a latent space with 1000 dimensions. The decoder consists
of a dense layer with 384 units with ReLU activation followed by an output layer
which matches the shape of the images and has a hyperbolic tangent activation
function. The reconstruction loss function used during training is taken to be
the mean squared error. The best parameters for the VAE model obtained after the
\texttt{hyperopt} procedure are shown in table~\ref{tab:vae-parameters}. In
figure~\ref{fig:events-vae} we show a random selection of preprocessed images
generated through the VAE. From a qualitative point of view the images appear
realistic on an event-by-event comparison however as highlighted in
section~\ref{sec:comparisons}, the VAE model does not reproduce the underlying
distribution accurately.

\begin{table}
  \begin{center}
    \phantom{x}\medskip
    \begin{tabular}{ccc}
      \toprule
      \textbf{Parameters} && \textbf{Value}\\
      \midrule
      $D$ units && 16\\ [3pt]
      $G$ units && 4\\ [3pt]
      $\alpha$ && 0.3\\ [3pt]
      Dropout && 0.15\\ [3pt]
      $D$ momentum && 0.7\\ [3pt]
      $G$ momentum && 0.7\\ [3pt]
      Minibatch discriminator && No\\
      \midrule
      Epochs && 300\\ [3pt]
      Batch size && 32\\ [3pt]
      Latent dimension && 800\\ [3pt]
      ZCA && Yes\\ [3pt]
      $n_{\rm avg}$ && 32\\ [3pt]
      Learning rate && $9.6 \cdot 10^{-5}$\\ [3pt]
      Decay $\beta_1$ && $2 \cdot 10^{-8}$\\ [3pt]
      $\rho$ && 0.9\\ [3pt]
      Optimiser && RMSprop\\
      \bottomrule
    \end{tabular}
    \caption{Final parameters for the WGAN-GP model.}
    \label{tab:wgan-gp-parameters}
  \end{center}
\end{table}

\begin{figure}
  \centering
  \includegraphics[width=0.5\textwidth]{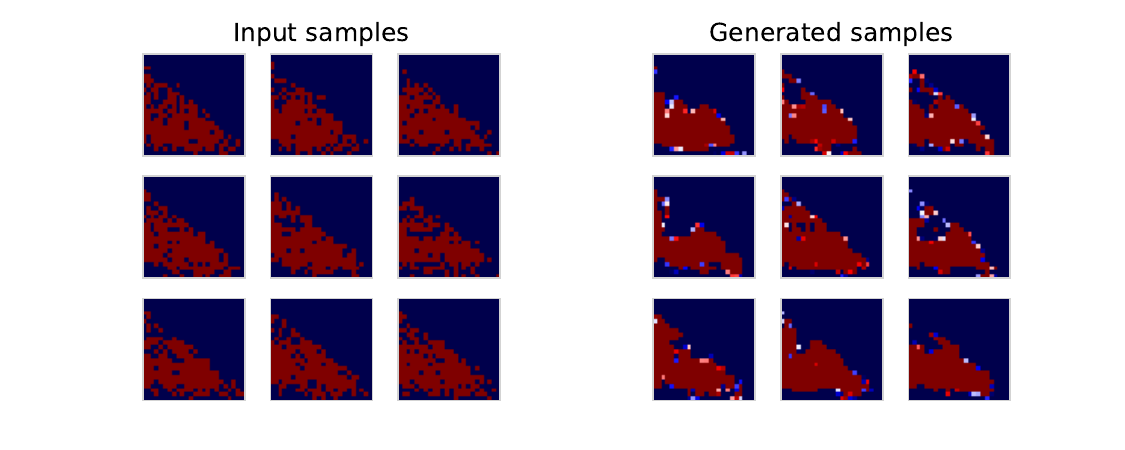}
  \caption{A random selection of preprocessed input images (left), and of
    images generated with the WGAN-GP model (right). Axis and colour schemes are the same of figure~\ref{fig:events}.}
  \label{fig:events-wgangp}
\end{figure}

Finally, the WGAN-GP consists in a generator and discriminator. The generator
architecture contains a dense layer with 1152 units with ReLU activation
function followed by three sequential two-dimensional convolutional layers with a
kernel size of 4 and respectively 32, 16 and 1 filters. Between these layers we
apply batch normalisation and ReLU activation function while the final layer has
a hyperbolic tangent activation function. On the other hand, the discriminator
is composed by 4 two-dimensional convolutional layers with a kernel size of 3
and respectively 16, 32, 64, 128 and 128 filters. We apply batch normalisation
for the last three layers and all of them LeakyReLU activation function with a
dropout layer. In table~\ref{tab:wgan-gp-parameters} we provide the best
parameters of the WGAN-GP model, always obtained through the \texttt{hyperopt}
scan procedure. In figure~\ref{fig:events-wgangp} we show a random selection of
preprocessed images generated through the WGAN-GP. Due to the convolutional
filters of this model the preprocessing differs slightly from the description in
section~\ref{sec:prob-generation} as we do not remove pixels outside the
kinematic range resulting in images with non zero background pixels. While
distributions presented in section~\ref{sec:comparisons} are in good agreement
with data, it is clear that for this WGAN-GP model the individual images look
different from the input data.

\bibliographystyle{epj}
\bibliography{generative}

\end{document}